\documentclass[onecolumn,showpacs,preprintnumbers]{revtex4}
\usepackage{mathrsfs}
\usepackage{graphicx}
\usepackage{dcolumn}
\usepackage{bm}
\usepackage{epsfig}
\usepackage{amsmath,amssymb}
\usepackage{multirow}

\begin{document}

\title{Mass spectra of double-bottom baryons }
\author{Zhen-Yu Li$^{1}$}
\email{zhenyvli@163.com }
\author{Guo-Liang Yu$^{2}$}
\email{yuguoliang2011@163.com }
\author{Zhi-Gang Wang$^{2}$ }
\email{zgwang@aliyun.com }
\author{Jian-Zhong Gu$^{3}$ }
\author{Hong-Tao Shen$^{4}$ }
\affiliation{$^1$ School of Physics and Electronic Science, Guizhou Education University, Guiyang 550018,
China\\$^2$ Department of Mathematics and Physics, North China Electric Power University, Baoding 071003,
China\\$^3$ China Institute of Atomic Energy, Beijing 102413,
China\\$^4$ Guangxi Key Laboratory of Nuclear Physics and Technology, Guangxi Normal University, Guilin 541006,
China}
\date{\today }

\begin{abstract}
Based on the relativistic quark model and the infinitesimally shifted Gaussian basis function method, we investigate the mass spectra of double bottom baryons systematically. In the $\rho$-mode which appears lower in energy than the other excited modes, we obtain the allowed quantum states and perform a systematic study of the mass spectra of the $\Xi_{bb}$  and $\Omega_{bb}$ families. We analyze the root mean square radii and quark radial probability density distributions to deeply understand the structure of the heavy baryons.  Meanwhile, the mass spectra allow us to successfully construct the Regge trajectories in the $(J,M^{2})$ plane. We also predict the masses of the ground states of double bottom baryons and discuss the differences between the structures of our spectra and those from other theoretical methods. At last, the shell structure of the double bottom baryon spectra is shown, from which one could get a bird's-eye view of the mass spectra.

\end{abstract}


\maketitle

\section*{I. Introduction}

\label{sec1}
The spectroscopy of doubly heavy baryons contains rich information of strong interactions and has become one of the hot topics in hadronic physics. In the past decades, research on the doubly heavy baryons
has developed rapidly in both experiments and theories.
The SELEX collaboration first reported the observation of the $\Xi_{cc}^{+}$ baryon in 2002~\cite{art01}. But the confirmation of the $\Xi_{cc}^{+}$ baryon was not yet completed and searching for the doubly heavy baryons came to a standstill.  Recently, the  $\Xi_{cc}^{++}$ baryon was observed by the LHCb collaboration~\cite{art02,art03,art04} and has been collected in the new PDG data~\cite{art05}, which is an important progress in the study of doubly heavy baryons. The efforts of searching for $\Xi_{bc}^{0}$~\cite{art06}, $\Omega_{bc}^{0}$~\cite{art07} and $\Xi_{bc}^{+}$~\cite{art08} baryons were reported one after another later on. Although the $\Xi_{bb}$ or $\Omega_{bb}$ has not been found experimentally, they are expected to be observed in the near future~\cite{art0801,art0802}.

On the other hand, the progress in experiment stimulates the theoretical studies on the spectroscopy of doubly heavy baryons, including the non-relativistic quark model~\cite{art09}, the effective field theory~\cite{art010}, the chiral perturbation theory~\cite{art011}, the MIT bag model~\cite{art012}, the effective QCD string theory~\cite{art013}, the QCD sum rules~\cite{art014,art0142,art0143}, the Bethe-Salpeter equation approach~\cite{art015,art016}, the di-quark picture in a
relativized quark model~\cite{art017}, the hyper-central constituent quark model~\cite{art018,art0182,art0183}, the lattice QCD~\cite{art019}, the chiral partner structure method~\cite{art020}, and the Regge phenomenology~\cite{art0201}. The theoretical studies mentioned above help us to understand the structure of double heavy baryons.

 Nevertheless, the theoretical predictions with different methods vary widely, probably being lack of experimental data. Table I lists the predicted masses of the ground states of double bottom baryons from different theoretical methods in the past three decades. As shown in Table I, the predicted masses of the ground states for the $\Xi_{bb}$ baryons range from 9800 MeV to 10340 MeV, and the maximal difference amounts to 540 MeV. In addition, there are problems in theory which need to be studied. For example, the structure of the mass spectra is not clear so far. So, it would be helpful to analyze the mass spectra of heavy baryons in a systematic way. Some theoretical efforts have been made indeed to solve these problems in this way~\cite{art013,art017,art018,art0182,art0183,art09,art025,art031,art056}.

In our previous papers~\cite{art021,art022}, we have analyzed the mass spectra of singly heavy baryons systematically, by using the relativistic quark model~\cite{art091,art09102} and the infinitesimally shifted Gaussian(ISG) function method~\cite{art092}. For singly heavy baryons, the excited state is confined to the $\lambda$-mode, which has a lower energy than the other modes.
 It turns out that most of the mass observed in the experiment can be reproduced well using our calculations. Because the $\lambda$-mode is associated with the heavy quark, it seems that heavy baryon excitation may be dominated by the excitation of heavy quarks in the heavy quark limit. If this conjecture holds, the excitation of the double heavy baryons should be dominated by the excitation of the two heavy quarks. According to this idea, in this paper, we will try to systematically analyze the mass spectra of double-bottom baryons, and further understand the structure of heavy baryons.

This paper is organized as follows. In
Sect.II, we briefly describe the methods used in the theoretical calculations; In Sect.III, we present the root mean square radii and the mass spectra of the doubly bottom baryons, analyze their quark radial probability density distributions, construct the Regge trajectories, and explore the spectral shell structure; And Sect.IV is reserved for our conclusions.

\begin{table*}[htbp]
\begin{ruledtabular}\caption{Predicted masses (in MeV) of the ground states of the double bottom baryons in the references. The superscript * refers to the $J=3/2$ baryons.}
\begin{tabular}{ c c c c c | c c c c c  }
$m(\Xi_{bb})$ & $m(\Xi_{bb}^{*})$ & $m(\Omega_{bb})$ & $m(\Omega_{bb}^{*})$ & year & $m(\Xi_{bb})$ & $m(\Xi_{bb}^{*})$ & $m(\Omega_{bb})$ & $m(\Omega_{bb}^{*})$ & year \\ \hline

        10322  & 10355 & 10500 & 10533 & 2022~\cite{art09} & 10143  & 10178 & 10273 & 10308 & 2014~\cite{art034}\\
        10221  & 10261 & -     & -     & 2022~\cite{art0183} & 10162  & 10184 & -     & -     & 2014~\cite{art035}\\
        10171  & 10195 & 10266 & 10291 & 2022~\cite{art025} & 10322  & 10352 & -     & -     & 2012~\cite{art036}\\
        10235  & -     & 10299 & -     & 2021~\cite{art011} & 9800   & 9890  & 9890  & 9930  & 2012~\cite{art037}\\
        10311  & 10360 & 10408 & 10451 & 2021~\cite{art012} & 10090  & -     & 10185 & -     & 2011~\cite{art038}\\
        10120  & 10150 & -     & -     & 2021~\cite{art013} & 10170  & 10220 & 10320 & 10380 & 2010~\cite{art039}\\
        10230  & 10333 & 10350 & 10449 & 2021~\cite{art026} & 10185  & 10216 & 10271 & 10289 & 2009~\cite{art040}\\
        10210  & 10221 & 10319 & 10331 & 2020~\cite{art027} & 10202  & 10237 & 10359 & 10389 & 2008~\cite{art042}\\
        10091  & 10103 & 10190 & 10203 & 2020~\cite{art028} & 10189  & 10218 & 10293 & 10321 & 2008~\cite{art043}\\
        10182  & 10214 & 10276 & 10309 & 2019~\cite{art015} & 10340  & 10367 & 10454 & 10486 & 2008~\cite{art045}\\
        10169  & 10189 & 10259 & 10268 & 2018~\cite{art029} & 10130  & 10144 & 10422 & 10432 & 2008~\cite{art046}\\
        -      & -     & 10208 & -     & 2018~\cite{art030} & 9780   & 10350 & 9850  & 10280  & 2008~\cite{art047}\\
        10220  & 10270 & 10330 & 10370 & 2018~\cite{art0142} &10062  & 10101 & 10208 & 10244 & 2008~\cite{art048}\\
        10250  & 10270 & 10340 & 10350 & 2018~\cite{art016} & 10197  & 10236 & 10260 & 10297 & 2007~\cite{art044}\\
        10317  & 10340 & -     & -     & 2017~\cite{art018} & 10100  & 10110 & 10280 & 10290 & 2004~\cite{art049}\\
        10138  & 10169 & 10230 & 10258 & 2017~\cite{art017} & 10202  & 10236 & 10359 & 10389 & 2002~\cite{art056}\\
        -      & -     & 10446 & 10467 & 2016~\cite{art0182} & 10090  & 10110 & 10210 & 10260 & 2002~\cite{art050}\\
        10199  & 10316 & 10320 & 10431 & 2015~\cite{art0201} & 10093  & 10133 & 10180 & 10200 & 2000~\cite{art051}\\
        10314  & 10339 & 10447 & 10467 & 2015~\cite{art031} & 10230  & 10280 & 10320 & 10360 & 1997~\cite{art052}\\
        10267  & -     & 10356 & -     & 2015~\cite{art032} & 10198  & 10236 & -     & -     & 1996~\cite{art053}\\
        10334  & 10431 & 10397 & 10495 & 2014~\cite{art033} & 10340  & 10370 & 10370 & 10400 & 1995~\cite{art054}\\

\end{tabular}
\end{ruledtabular}
\end{table*}

\section*{II. Phenomenological methods adopted in this work}

In this work, we use the relativistic quark model and the ISG method to study double bottom baryons. The detailed discussions about these two methods can be found in references~\cite{art091,art09102,art092,art021,art022}.
 The doubly heavy baryon is a three-quark system, which is commonly studied in the Jacobi coordinates. As shown in Fig.1, there are three channels of the Jacobi coordinates for the three-quark system.
The corresponding Jacobi coordinates are defined as
\begin{eqnarray}
&\boldsymbol\rho_{i}=\textbf{r}_{j}-\textbf{r}_{k}, \\
&\boldsymbol\lambda_{i}=\textbf{r}_{i}-\frac{m_{j}\textbf{r}_{j}+m_{k}\textbf{r}_{k}}{m_{j}+m_{k}},
\end{eqnarray}
where $i$, $j$, $k$ = 1, 2, 3 (or replace their positions in turn). $\textbf{r}_{i}$ and $m_{i}$ denote the position vector and the mass of the $i$th quark, respectively.
\begin{figure}[htbp]
\begin{center}
\includegraphics[width=0.8\textwidth]{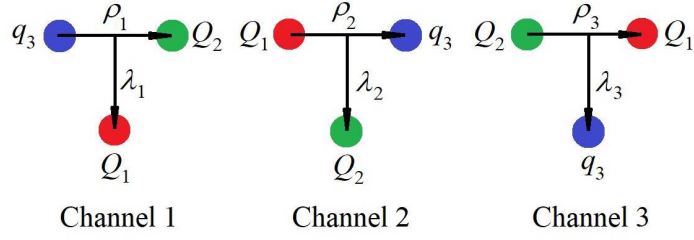}
\end{center}
\caption{(Color online)Jacobi coordinates for the three-body system. We denote the light quark as the 3rd particle in the case of doubly heavy baryons. }
\end{figure}

A double bottom baryon  is  commonly considered as a system composed of a heavy di-quark and a light quark ~\cite{art015,art017,art023}. Accordingly,
 the calculations in this work are based on channel 3. In this case, the 3rd quark is just the light quark. $\textbf{\emph{l}}_{\rho3}$ (denoted in short as $\textbf{\emph{l}}_{\rho}$) is defined as the orbital angular momentum between the two bottom quarks, and  $\textbf{\emph{l}}_{\lambda3}$ (denoted in short as $\textbf{\emph{l}}_{\lambda}$) represents the one between the bottom-quark pair and the light quark.

For a definite state in theory, the spatial wave function and the spin function are written as follow,
\begin{eqnarray}
\begin{aligned}
|l_{\rho} \ l_{\lambda} \ L \ s\ j\ J\ M_{J}\rangle
&=\{[(|l_{\rho}\ m_{\rho} \rangle |l_{\lambda}\ m_{\lambda} \rangle)_{L}\times(|s_{1}\ m_{s_{1}} \rangle|s_{2}\ m_{s_{2}} \rangle)_{s}]_{j}\times|s_{3}\ m_{s_{3}} \rangle \}_{J M_{J}}.
\end{aligned}
\end{eqnarray}
$l_{\rho}$, $l_{\lambda}$, $L$, $s$, $j$, $J$ and $M_{J}$  are quantum numbers which characterize a given state. Because the total wave function must be antisymmetric and the flavor function of the double bottom quark subsystem ($bb$) is symmetric, the total spin $s$  and orbital quantum number $l_{\rho}$ of ($bb$) should meet the condition $(-1)^{s+l_{\rho}}=-1$.

\section*{III. Numerical results and discussions}

\subsection*{3.1  $\rho$-mode  }

  $nL(J^{P})$ is usually used to describe a baryon state in experiment, where the orbital quantum numbers $\textbf{\emph{l}}_{\rho}$ and $\textbf{\emph{l}}_{\lambda}$ are unmeasurable. For $L\neq0$, there usually exist three ($l_{\rho}$, $l_{\lambda}$) modes under the condition $\textbf{L}=\textbf{\emph{l}}_{\rho}+\textbf{\emph{l}}_{\lambda}$: (1) The $\rho$-mode with $l_{\rho}\neq0$ and $l_{\lambda}=0$; (2) The $\lambda$-mode with $l_{\rho}=0$ and $l_{\lambda}\neq0$; (3) The $\lambda$-$\rho$ mixing mode with $l_{\rho}\neq0$ and $l_{\lambda}\neq0$. These ($l_{\rho}$, $l_{\lambda}$) modes should generally be mixed.

 We first investigate the excitation energy for different modes.
 As an example, the excitation energies of the $1D(\frac{3}{2}^{+}, \frac{5}{2}^{+})_{j=2}$ states as functions of $m_{Q}$ are investigated. As shown in Fig.2, the excitation energies of three modes are crossed in the range of 0.2-0.9 GeV, where the heavy quark limit is invalid and the mixing between these modes can not be neglected.
 However, they are separated when $m_{Q}$ increases from 0.9 GeV to 5.0 GeV. And the $\rho$-mode appears lower in energy than the other two modes in the heavy quark limit for double bottom baryons, which is different from that of the singly heavy baryons~\cite{art021,art022}. This confirms the conjecture mentioned in the introduction that the heavy quark excitation is dominant. As a reasonable approximation, we only study the $\rho$-mode in this work.

  \begin{figure}[htbp]
\begin{center}
\includegraphics[width=0.6\textwidth]{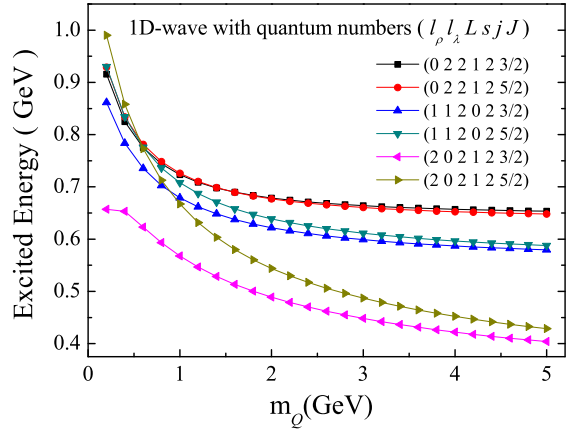}
\end{center}
\caption{(Color online)The dependence of excitation energies on $m_{Q}$ for different modes of $\Xi_{QQ}$, where $m_{1}=m_{2}=m_{Q}$ and $m_{3}=m_{u(d)}$. The excitation energies are measured from the ground state with quantum numbers ($l_{\rho}\ l_{\lambda}\ L\ s\ j\ J$)=(0 0 0 1 1 1/2). }
\end{figure}

\subsection*{3.2 Mass spectra, root mean square radii and quark radial probability density distributions}

In the $\rho$ mode, the root mean square radii, quark radial probability density distributions and mass spectra of the double bottom baryons with quantum numbers up to $n=4$ and $L=4$ are presented in Tables II-III and Figs.3-4.

 Through the analysis of these calculated results, some general features of the mass spectra are summarized as follows:
 (1) There are a total of 18 quantum states with $L\leq4$ in the $\Xi_{bb}$ or $\Omega_{bb}$ family.
 (2) For the same $L$, the mass splitting  becomes larger with the increase of $j$. For example, Table I shows the mass differences (splittings) of the $1D$ doublets with $j=1,2,3$ are 15 MeV, 25 MeV and 35 MeV, respectively. This is apparently different from that of the singly heavy baryons~\cite{art021,art022}.
 (3) The mass difference between the two adjacent radial excited states gradually decreases with increasing $n$.

On the other hand, the calculated root mean square radii and quark radial probability density distributions carry important information. For a three-quark system, the radial probability densities $\omega(r_{\rho})$ and $\omega(r_{\lambda})$ can be defined as follows,
\begin{eqnarray}
\begin{aligned}
& \omega(r_{\rho})=\int |\Psi(\textbf{r}_{\rho},\textbf{r}_{\lambda})|^{2}\mathrm{d}\textbf{r}_{\lambda}\mathrm{d}\Omega_{\rho},\\
&  \omega(r_{\lambda})=\int |\Psi(\textbf{r}_{\rho},\textbf{r}_{\lambda})|^{2}\mathrm{d}\textbf{r}_{\rho}\mathrm{d}\Omega_{\lambda},
\end{aligned}
\end{eqnarray}
where $\Omega_{\rho}$ and $\Omega_{\lambda}$ are the solid angles spanned by vectors $\textbf{r}_{\rho}$ and $\textbf{r}_{\lambda}$, respectively. From Figs.3-4 and Tables II-III, one can find some interesting properties.
(1) For a ground state, the $\langle r_{\rho}^{2}\rangle ^{1/2}$ value is much smaller than the $\langle r_{\lambda}^{2}\rangle ^{1/2}$ value. This indicates that the two bottom quarks are bonded very tightly.
(2) As shown by the two types of the root mean square radii in Table II, the main radial excitation happens in the  $r_{\rho}$ and $r_{\lambda}$ spaces in turn with $n=2,3,4$. This feature can be also seen in the quark radial probability density distributions in Figs.3-4.
(3) As shown in Tables.II-III, for the orbital excited  states with fixed $n$, the difference of their $\langle r_{\lambda}^{2}\rangle ^{1/2}$ values is relatively small. While their $\langle r_{\rho}^{2}\rangle ^{1/2}$ values become larger with increasing $L$. In Figs.3 and 4, the radial probability density distribution of $r_{\lambda}^{2}\omega(r_{\lambda})$  (dash lines) changes a little with different $L$ values. But, the peak of the $r_{\rho}^{2}\omega(r_{\rho})$ (solid lines) is significantly shifted outward with increasing $L$.
(4) The shapes of the black (solid) lines in Fig.3 are almost as same as those in Fig.4. And, for the same state, the $\langle r_{\rho}^{2}\rangle ^{1/2}$ value in the $\Xi_{bb}$ family is almost the same as that in the $\Omega_{bb}$ family. This reflects the similarity in structure of the $\Xi_{bb}$ and $\Omega_{bb}$ baryons.

\begin{table*}[htbp]
\begin{ruledtabular}\caption{The root mean square radii (fm) and mass spectra (MeV) of the $\Xi_{bb}$ family.}
\begin{tabular}{c c c c c | c c c c c}
$l_{\rho}$  $l_{\lambda}$ $L$ $s$ $j$  &$nL$($J^{P}$) & $\langle r_{\rho}^{2}\rangle^{1/2}$ & $\langle r_{\lambda}^{2}\rangle^{1/2}$ & mass & $l_{\rho}$  $l_{\lambda}$ $L$ $s$ $j$  &$nL$($J^{P}$) & $\langle r_{\rho}^{2}\rangle^{1/2}$ & $\langle r_{\lambda}^{2}\rangle^{1/2}$ & mass \\ \hline
\multirow{4}{*}{0 0 0 1 1 }
        & $1S$($\frac{1}{2}^{+}$)  & 0.297 & 0.469 & 10192 & \multirow{4}{*}{2 0 2 1 3} & $1D$($\frac{7}{2}^{+}$) & 0.596  & 0.540 & 10629 \\
        & $2S$($\frac{1}{2}^{+}$)  & 0.585 & 0.522 & 10536 & ~ & $2D$($\frac{7}{2}^{+}$)  & 0.893 & 0.595 & 10863 \\
        & $3S$($\frac{1}{2}^{+}$)  & 0.312 & 0.839 & 10686 & ~ & $3D$($\frac{7}{2}^{+}$)  & 0.619 & 0.899 & 11094  \\
        & $4S$($\frac{1}{2}^{+}$)  & 0.757 & 0.555 & 10796 & ~ & $4D$($\frac{7}{2}^{+}$)  & 0.999 & 0.617 & 11159    \\ \hline
\multirow{4}{*}{0 0 0 1 1}
        & $1S$($\frac{3}{2}^{+}$)  & 0.299 & 0.483& 10211 & \multirow{4}{*}{3 0 3 0 3} & $1F$($\frac{5}{2}^{-}$) & 0.702 & 0.534 & 10741 \\
        & $2S$($\frac{3}{2}^{+}$)  & 0.589 & 0.534& 10552 & ~ & $2F$($\frac{5}{2}^{-}$)  & 0.997 & 0.589 & 10955 \\
        & $3S$($\frac{3}{2}^{+}$)  & 0.312 & 0.847& 10696 & ~ & $3F$($\frac{5}{2}^{-}$)  & 0.737 & 0.901 & 11211  \\
        & $4S$($\frac{3}{2}^{+}$)  & 0.760 & 0.566& 10810 & ~ & $4F$($\frac{5}{2}^{-}$)  & 1.253 & 0.648 & 11295    \\ \hline
\multirow{4}{*}{1 0 1 0 1 }
        & $1P$($\frac{1}{2}^{-}$)  & 0.456 & 0.496& 10428 & \multirow{4}{*}{3 0 3 0 3} & $1F$($\frac{7}{2}^{-}$) & 0.711  & 0.562 & 10773\\
        & $2P$($\frac{1}{2}^{-}$)  & 0.746 & 0.550& 10701 & ~ & $2F$($\frac{7}{2}^{-}$)  & 0.996 & 0.614 & 10981 \\
        & $3P$($\frac{1}{2}^{-}$)  & 0.486 & 0.857& 10912 & ~ & $3F$($\frac{7}{2}^{-}$)  & 0.740 & 0.923 & 11231 \\
        & $4P$($\frac{1}{2}^{-}$)  & 0.809 & 0.574& 10959 & ~ & $4F$($\frac{7}{2}^{-}$)  & 1.346 & 0.687 & 11322    \\ \hline
\multirow{4}{*}{1 0 1 0 1 }
        & $1P$($\frac{3}{2}^{-}$)  & 0.459 & 0.508& 10445 & \multirow{4}{*}{4 0 4 1 3} & $1G$($\frac{5}{2}^{+}$) & 0.812 & 0.555 & 10876 \\
        & $2P$($\frac{3}{2}^{-}$)  & 0.750 & 0.561& 10715 & ~ & $2G$($\frac{5}{2}^{+}$)  & 1.078 & 0.604 & 11067 \\
        & $3P$($\frac{3}{2}^{-}$)  & 0.484 & 0.867& 10922 & ~ & $3G$($\frac{5}{2}^{+}$)  & 0.861 & 0.923 & 11336 \\
        & $4P$($\frac{3}{2}^{-}$)  & 0.818 & 0.583& 10973 & ~ & $4G$($\frac{5}{2}^{+}$)  & 1.653 & 0.742 & 11441    \\ \hline
\multirow{4}{*}{2 0 2 1 1}
        & $1D$($\frac{1}{2}^{+}$)  & 0.584 & 0.519& 10600& \multirow{4}{*}{4 0 4 1 3} & $1G$($\frac{7}{2}^{+}$) & 0.824 & 0.582 & 10905 \\
        & $2D$($\frac{1}{2}^{+}$)  & 0.886 & 0.576& 10839 & ~ & $2G$($\frac{7}{2}^{+}$)  & 1.073 & 0.627 & 11092 \\
        & $3D$($\frac{1}{2}^{+}$)  & 0.614 & 0.880& 11074 & ~ & $3G$($\frac{7}{2}^{+}$)  & 0.865 & 0.945 & 11356 \\
        & $4D$($\frac{1}{2}^{+}$)  & 0.921 & 0.592& 11127 & ~ & $4G$($\frac{7}{2}^{+}$)  & 1.709 & 0.769 & 11461 \\ \hline
\multirow{4}{*}{2 0 2 1 1}
        & $1D$($\frac{3}{2}^{+}$)  & 0.587 & 0.531& 10615 & \multirow{4}{*}{4 0 4 1 4} & $1G$($\frac{7}{2}^{+}$) & 0.812 & 0.551 & 10872 \\
        & $2D$($\frac{3}{2}^{+}$)  & 0.888 & 0.587& 10851 & ~ & $2G$($\frac{7}{2}^{+}$)  & 1.078 & 0.600 & 11063 \\
        & $3D$($\frac{3}{2}^{+}$)  & 0.614 & 0.890& 11084 & ~ & $3G$($\frac{7}{2}^{+}$)  & 0.862 & 0.921 & 11334 \\
        & $4D$($\frac{3}{2}^{+}$)  & 0.943 & 0.603& 11141 & ~ & $4G$($\frac{7}{2}^{+}$)  & 1.655 & 0.740 & 11439    \\ \hline
\multirow{4}{*}{2 0 2 1 2}
        & $1D$($\frac{3}{2}^{+}$)  & 0.585 & 0.516& 10596 & \multirow{4}{*}{4 0 4 1 4 } & $1G$($\frac{9}{2}^{+}$) & 0.827  & 0.586 & 10909 \\
        & $2D$($\frac{3}{2}^{+}$) & 0.887 & 0.572 & 10836 & ~ & $2G$($\frac{9}{2}^{+}$)  & 1.071 & 0.630 & 11096 \\
        & $3D$($\frac{3}{2}^{+}$)  & 0.616 & 0.878& 11072 & ~ & $3G$($\frac{9}{2}^{+}$)  & 0.867 & 0.948 & 11359 \\
        & $4D$($\frac{3}{2}^{+}$)  & 0.927 & 0.590& 11126 & ~ & $4G$($\frac{9}{2}^{+}$)  & 1.723 & 0.773 & 11464    \\ \hline
\multirow{4}{*}{2 0 2 1 2 }
        & $1D$($\frac{5}{2}^{+}$)  & 0.591 & 0.536& 10621 & \multirow{4}{*}{4 0 4 1 5} & $1G$($\frac{9}{2}^{+}$) &0.813 &0.547 & 10868 \\
        & $2D$($\frac{5}{2}^{+}$)  & 0.890 & 0.591& 10856 & ~ & $2G$($\frac{9}{2}^{+}$)  & 1.078 & 0.597 & 11060 \\
        & $3D$($\frac{5}{2}^{+}$) & 0.615 & 0.894 & 11088 & ~ & $3G$($\frac{9}{2}^{+}$)  & 0.865 & 0.918 & 11331 \\
        & $4D$($\frac{5}{2}^{+}$)  & 0.964 & 0.609& 11149 & ~ & $4G$($\frac{9}{2}^{+}$)  & 1.663 & 0.738 & 11436 \\ \hline
\multirow{4}{*}{2 0 2 1 3 }
        & $1D$($\frac{5}{2}^{+}$)  & 0.588 & 0.512& 10594 & \multirow{4}{*}{4 0 4 1 5} & $1G$($\frac{11}{2}^{+}$) &0.831 & 0.590 &10914\\
        & $2D$($\frac{5}{2}^{+}$)  & 0.889 & 0.569& 10834 & ~ & $2G$($\frac{11}{2}^{+}$)  & 1.070 & 0.633 & 11100 \\
        & $3D$($\frac{5}{2}^{+}$)  & 0.619 & 0.876& 11072 & ~ & $3G$($\frac{11}{2}^{+}$)  & 0.871 & 0.952 & 11363 \\
        & $4D$($\frac{5}{2}^{+}$)  & 0.942 & 0.588& 11127 & ~ & $4G$($\frac{11}{2}^{+}$)  & 1.741 & 0.777 & 11467 \\
\end{tabular}
\end{ruledtabular}
\end{table*}

\begin{table*}[htbp]
\begin{ruledtabular}\caption{The root mean square radii (fm) and mass spectra (MeV) of the $\Omega_{bb}$ family.}
\begin{tabular}{c c c c c | c c c c c}
$l_{\rho}$  $l_{\lambda}$ $L$ $s$ $j$  &$nL$($J^{P}$) & $\langle r_{\rho}^{2}\rangle^{1/2}$ & $\langle r_{\lambda}^{2}\rangle^{1/2}$ & mass & $l_{\rho}$  $l_{\lambda}$ $L$ $s$ $j$  &$nL$($J^{P}$) & $\langle r_{\rho}^{2}\rangle^{1/2}$ & $\langle r_{\lambda}^{2}\rangle^{1/2}$ & mass \\ \hline
\multirow{4}{*}{0 0 0 1 1 }
        & $1S$($\frac{1}{2}^{+}$)  & 0.293 & 0.426 & 10285 & \multirow{4}{*}{2 0 2 1 3} & $1D$($\frac{7}{2}^{+}$) & 0.587  & 0.496 & 10731 \\
        & $2S$($\frac{1}{2}^{+}$)  & 0.573 & 0.483 & 10638 & ~ & $2D$($\frac{7}{2}^{+}$)  & 0.887 & 0.552 & 10972 \\
        & $3S$($\frac{1}{2}^{+}$)  & 0.314 & 0.782 & 10777 & ~ & $3D$($\frac{7}{2}^{+}$)  & 0.615 & 0.844 & 11190  \\
        & $4S$($\frac{1}{2}^{+}$)  & 0.750 & 0.517 & 10902 & ~ & $4D$($\frac{7}{2}^{+}$)  & 0.923 & 0.568 & 11264    \\ \hline
\multirow{4}{*}{0 0 0 1 1}
        & $1S$($\frac{3}{2}^{+}$)  & 0.294 & 0.438& 10303 & \multirow{4}{*}{3 0 3 0 3} & $1F$($\frac{5}{2}^{-}$) & 0.691 & 0.494 & 10851 \\
        & $2S$($\frac{3}{2}^{+}$)  & 0.577 & 0.493& 10653 & ~ & $2F$($\frac{5}{2}^{-}$)  & 0.997 & 0.551 & 11071 \\
        & $3S$($\frac{3}{2}^{+}$)  & 0.314 & 0.790& 10787 & ~ & $3F$($\frac{5}{2}^{-}$)  & 0.732 & 0.847 & 11310  \\
        & $4S$($\frac{3}{2}^{+}$)  & 0.752 & 0.526& 10915 & ~ & $4F$($\frac{5}{2}^{-}$)  & 1.114 & 0.601 & 11407    \\ \hline
\multirow{4}{*}{1 0 1 0 1 }
        & $1P$($\frac{1}{2}^{-}$)  & 0.449 & 0.454& 10528 & \multirow{4}{*}{3 0 3 0 3} & $1F$($\frac{7}{2}^{-}$) & 0.700  & 0.517 & 10879\\
        & $2P$($\frac{1}{2}^{-}$)  & 0.734 & 0.509& 10809 & ~ & $2F$($\frac{7}{2}^{-}$)  & 0.997 & 0.572 & 11094 \\
        & $3P$($\frac{1}{2}^{-}$)  & 0.483 & 0.803& 11007 & ~ & $3F$($\frac{7}{2}^{-}$)  & 0.735 & 0.868 & 11329 \\
        & $4P$($\frac{1}{2}^{-}$)  & 0.792 & 0.533& 11064 & ~ & $4F$($\frac{7}{2}^{-}$)  & 1.198 & 0.636 & 11432    \\ \hline
\multirow{4}{*}{1 0 1 0 1 }
        & $1P$($\frac{3}{2}^{-}$)  & 0.451 & 0.465& 10543 & \multirow{4}{*}{4 0 4 1 3} & $1G$($\frac{5}{2}^{+}$) & 0.799 & 0.514 & 10990 \\
        & $2P$($\frac{3}{2}^{-}$)  & 0.738 & 0.519& 10821 & ~ & $2G$($\frac{5}{2}^{+}$)  & 1.083 & 0.566 & 11186 \\
        & $3P$($\frac{3}{2}^{-}$)  & 0.482 & 0.813& 11016 & ~ & $3G$($\frac{5}{2}^{+}$)  & 0.854 & 0.870 & 11437 \\
        & $4P$($\frac{3}{2}^{-}$)  & 0.797 & 0.541& 11077 & ~ & $4G$($\frac{5}{2}^{+}$)  & 1.456 & 0.728 & 11560    \\ \hline
\multirow{4}{*}{2 0 2 1 1}
        & $1D$($\frac{1}{2}^{+}$)  & 0.575 & 0.478& 10704& \multirow{4}{*}{4 0 4 1 3} & $1G$($\frac{7}{2}^{+}$) & 0.809 & 0.537 & 11014 \\
        & $2D$($\frac{1}{2}^{+}$)  & 0.879 & 0.535& 10951 & ~ & $2G$($\frac{7}{2}^{+}$)  & 1.079 & 0.586 & 11207 \\
        & $3D$($\frac{1}{2}^{+}$)  & 0.609 & 0.826& 11171 & ~ & $3G$($\frac{7}{2}^{+}$)  & 0.858 & 0.890 & 11455 \\
        & $4D$($\frac{1}{2}^{+}$)  & 0.865 & 0.548& 11233 & ~ & $4G$($\frac{7}{2}^{+}$)  & 1.530 & 0.751 & 11580 \\ \hline
\multirow{4}{*}{2 0 2 1 1}
        & $1D$($\frac{3}{2}^{+}$)  & 0.578 & 0.489& 10718 & \multirow{4}{*}{4 0 4 1 4} & $1G$($\frac{7}{2}^{+}$) & 0.799 & 0.511 & 10986 \\
        & $2D$($\frac{3}{2}^{+}$)  & 0.881 & 0.545& 10962 & ~ & $2G$($\frac{7}{2}^{+}$)  & 1.083 & 0.563 & 11183 \\
        & $3D$($\frac{3}{2}^{+}$)  & 0.609 & 0.836& 11180 & ~ & $3G$($\frac{7}{2}^{+}$)  & 0.856 & 0.867 & 11434 \\
        & $4D$($\frac{3}{2}^{+}$)  & 0.881 & 0.558& 11246 & ~ & $4G$($\frac{7}{2}^{+}$)  & 1.457 & 0.726 & 11558    \\ \hline
\multirow{4}{*}{2 0 2 1 2}
        & $1D$($\frac{3}{2}^{+}$)  & 0.576 & 0.475& 10702 & \multirow{4}{*}{4 0 4 1 4 } & $1G$($\frac{9}{2}^{+}$) & 0.812  & 0.541 & 11018 \\
        & $2D$($\frac{3}{2}^{+}$) & 0.880 & 0.533 & 10948 & ~ & $2G$($\frac{9}{2}^{+}$)  & 1.078 & 0.588 & 11210 \\
        & $3D$($\frac{3}{2}^{+}$)  & 0.611 & 0.824& 11170 & ~ & $3G$($\frac{9}{2}^{+}$)  & 0.861 & 0.893 & 11457 \\
        & $4D$($\frac{3}{2}^{+}$)  & 0.869 & 0.546& 11232 & ~ & $4G$($\frac{9}{2}^{+}$)  & 1.550 & 0.755 & 11582    \\ \hline
\multirow{4}{*}{2 0 2 1 2 }
        & $1D$($\frac{5}{2}^{+}$)  & 0.582 & 0.492& 10724 & \multirow{4}{*}{4 0 4 1 5} & $1G$($\frac{9}{2}^{+}$) &0.800 &0.508 & 10983 \\
        & $2D$($\frac{5}{2}^{+}$)  & 0.884 & 0.548& 10966 & ~ & $2G$($\frac{9}{2}^{+}$)  & 1.083 & 0.560 & 11180 \\
        & $3D$($\frac{5}{2}^{+}$) & 0.611 & 0.839 & 11184 & ~ & $3G$($\frac{9}{2}^{+}$)  & 0.858 & 0.865 & 11432 \\
        & $4D$($\frac{5}{2}^{+}$)  & 0.896 & 0.562& 11253 & ~ & $4G$($\frac{9}{2}^{+}$)  & 1.466 & 0.726 & 11557 \\ \hline
\multirow{4}{*}{2 0 2 1 3 }
        & $1D$($\frac{5}{2}^{+}$)  & 0.579 & 0.472& 10700 & \multirow{4}{*}{4 0 4 1 5} & $1G$($\frac{11}{2}^{+}$) &0.816 & 0.544 &11022\\
        & $2D$($\frac{5}{2}^{+}$)  & 0.883 & 0.530& 10947 & ~ & $2G$($\frac{11}{2}^{+}$)  & 1.076 & 0.591 & 11213 \\
        & $3D$($\frac{5}{2}^{+}$)  & 0.615 & 0.822& 11170 & ~ & $3G$($\frac{11}{2}^{+}$)  & 0.865 & 0.897 & 11461 \\
        & $4D$($\frac{5}{2}^{+}$)  & 0.881 & 0.545& 11235 & ~ & $4G$($\frac{11}{2}^{+}$)  & 1.576 & 0.759 & 11586 \\
\end{tabular}
\end{ruledtabular}
\end{table*}

\begin{figure}[htbp]
\begin{center}
\includegraphics[width=1.0\textwidth]{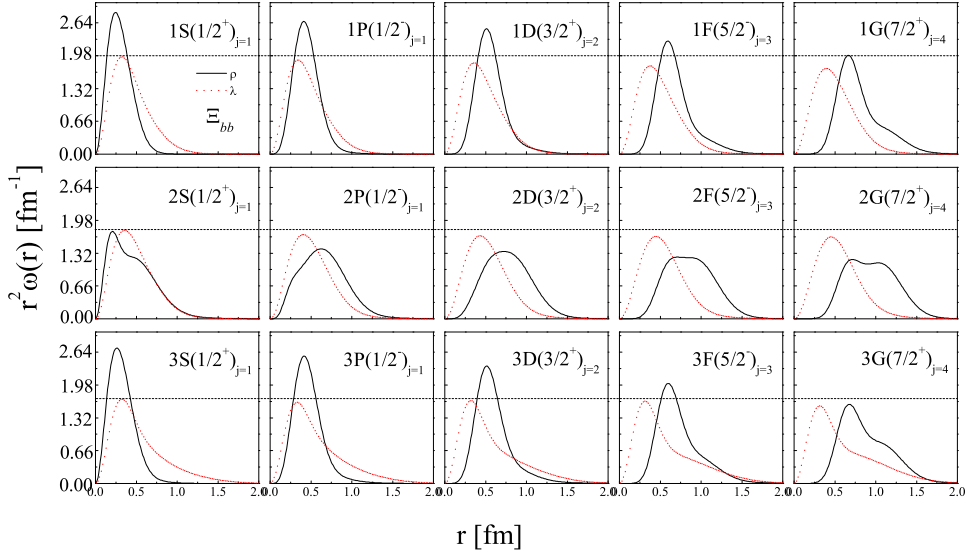}
\end{center}
\caption{(Color online)Radial probability density distributions for some $nL$ states of the $\Xi_{bb}$ family. The solid line denotes the probability density with $r_{\rho}$, and the dash line denotes the one with $r_{\lambda}$.}
\end{figure}

\begin{figure}[htbp]
\begin{center}
\includegraphics[width=1.0\textwidth]{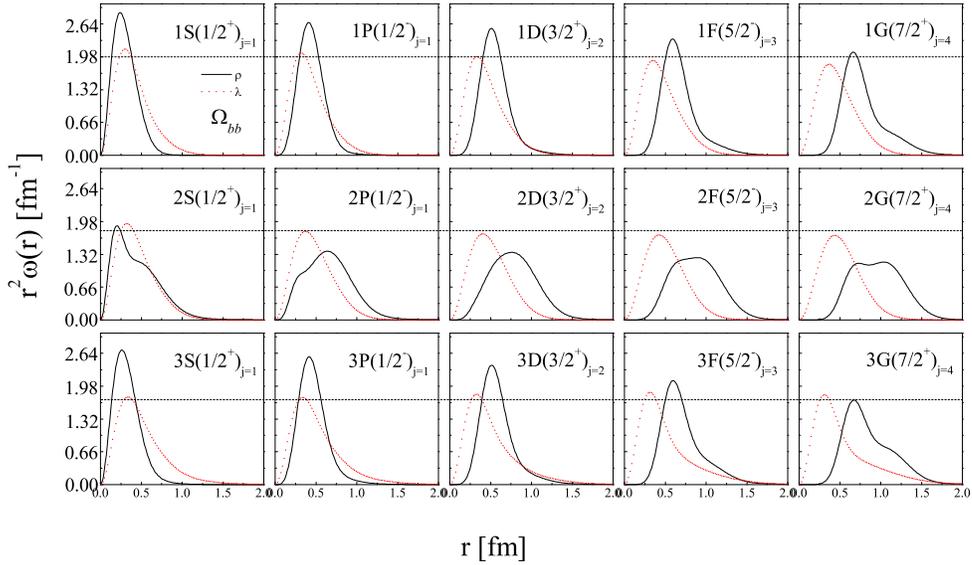}
\end{center}
\caption{(Color online)Same as Fig.3, but for the $\Omega_{bb}$ family.}
\end{figure}

\subsection*{3.3 Regge trajectories}

 For singly heavy baryons, we have successfully constructed the Regge trajectories\cite{art021,art022,art503,art504,art501,art502}, where the mass spectra were obtained in the $\lambda$-mode. In this work, we study the Regge trajectories of double bottom baryons in the $\rho$-mode.
 We use the following definition for the $(J, M^{2})$ Regge trajectories,
 \begin{eqnarray}
M^{2}=\alpha J+ \beta,
\end{eqnarray}
where $\alpha$ and $\beta$ are the slope and intercept. In Figs.5-6, we plot the Regge trajectories in the $(J, M^{2})$
plane with our calculated mass spectra. The four lines in each figure correspond to the radial quantum number $n$= 1, 2, 3, 4, respectively.
 The fitted slopes and intercepts of the Regge trajectories are given in Table IV.

  As shown in Fig.5, the group with natural parity (NP) $(-1)^{J-1/2}$ is composed of $S(\frac{1}{2}^{+})_{j=1}$, $P(\frac{3}{2}^{-})_{j=1}$, $D(\frac{5}{2}^{+})_{j=2}$, $F(\frac{7}{2}^{-})_{j=3}$ and $G(\frac{9}{2}^{+})_{j=4}$ states. The group with the unnatural parity (UP) $(-1)^{J+1/2}$ is composed of $P(\frac{1}{2}^{-})_{j=1}$, $D(\frac{3}{2}^{+})_{j=2}$, $F(\frac{5}{2}^{-})_{j=3}$ and $G(\frac{7}{2}^{+})_{j=4}$ states.
  In fact, there are 18 members of the $\Xi_{bb}$ family in this work. The remaining 9 states can also be put into these lines, because their mass values are very near those states with the same $L(J^{P})$. The situation is similar for the $\Omega_{bb}$ family as shown in Fig.6.

It is shown that the linear trajectories appear clearly in the $(J, M^{2})$ plane. Most of the data points fall on the trajectory lines. This indicates that the Regge trajectories have a strong universality. However, these lines are in whole not equidistant, which is different from those in references~\cite{art018,art0182,art0183}.

On the other hand, the linear trajectories in the $(n, M^{2})$ plane can not be constructed from our predicted masses, the reason is similar to that of the singly heavy baryons~\cite{art021,art022}. But, the linear Regge trajectories in the $(n, M^{2})$ plane for double bottom baryons are obtained within a hyper-central constituent quark model~\cite{art0182}. If the experiment in the near future touches the sub-shells of $n=$ 3 and 4, it might be a good time to check the $(n, M^{2})$ Regge trajectories.

\begin{figure}[htbp]
\begin{center}
\includegraphics[width=0.6\textwidth]{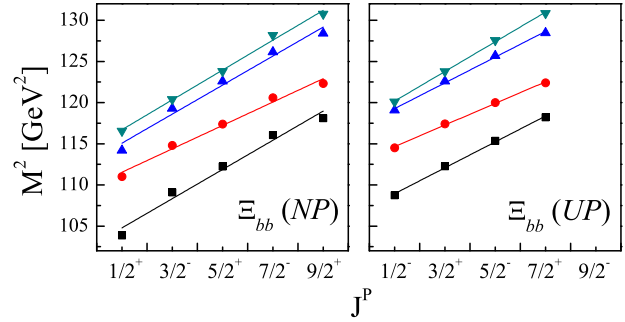}
\end{center}
\caption{(Color online)$(J, M^{2})$ Regge trajectories for the $\Xi_{bb}$ family and $M^{2}$ is in GeV$^{2}$. The $NP$ denotes the natural parity, and the $UP$ denotes the unnatural parity. }
\end{figure}

\begin{figure}[htbp]
\begin{center}
\includegraphics[width=0.6\textwidth]{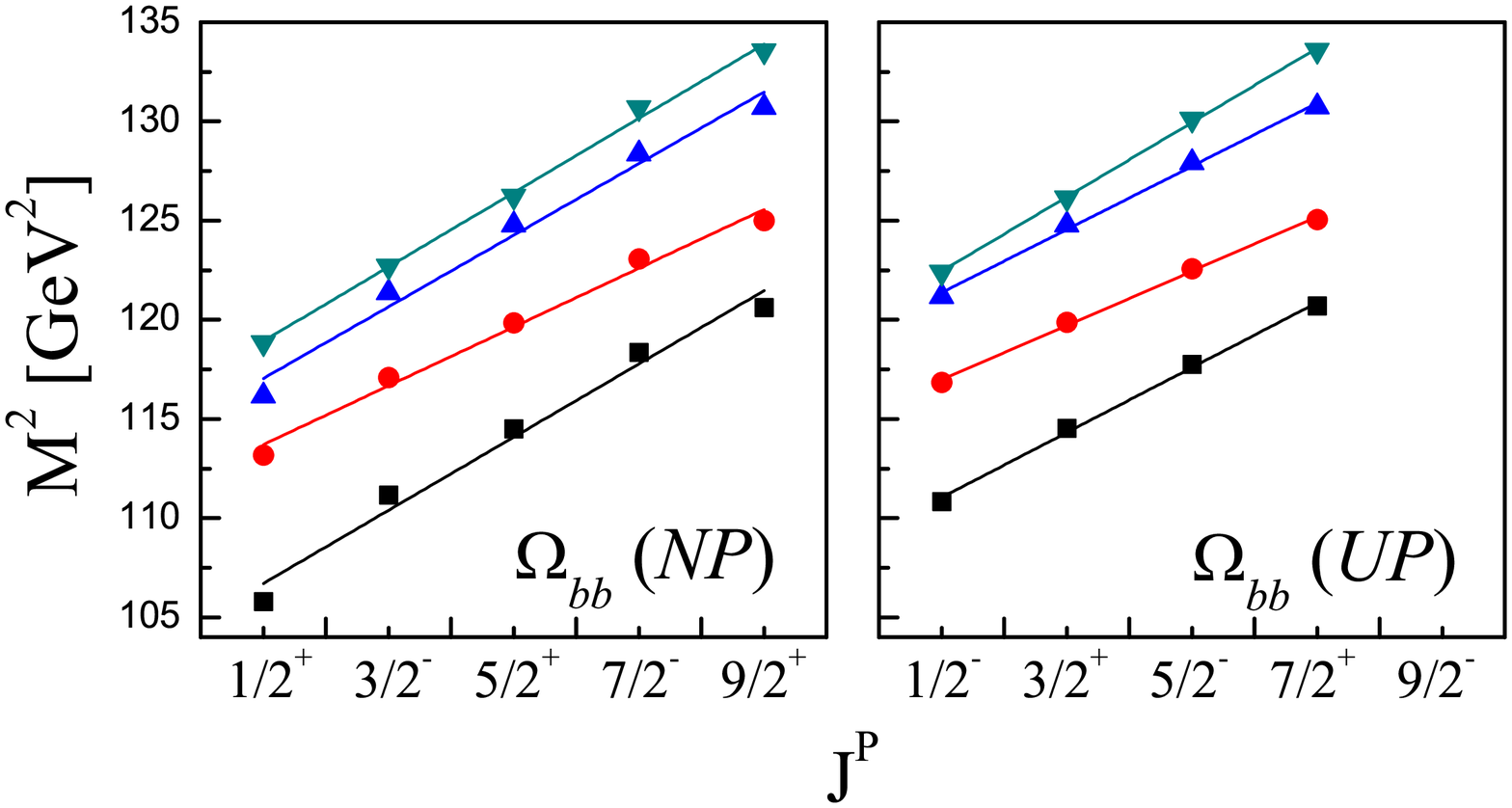}
\end{center}
\caption{(Color online)Same as Fig.5, but for the $\Omega_{bb}$ family.}
\end{figure}

\begin{table*}[htbp]
\begin{ruledtabular}\caption{Fitted values of the slope and intercept of the Regge trajectories for the $\Xi_{c}$ and $\Xi_{c}^{'}$ families.}
\begin{tabular}{c |c c| c c }
\multirow{2}{*}{Trajectory} & \multicolumn{2}{c|}{$\Xi_{bb}$}   &\multicolumn{2}{c}{$\Omega_{bb}$}   \\
& \ $\alpha$(GeV$^{2}$) & \ $\beta$(GeV$^{2}$) & \ $\alpha$(GeV$^{2}$) & \ $\beta$(GeV$^{2}$)   \\
\hline
$n=1(NP)$ &\ $3.547\pm0.301$ &\ $103.003\pm0.864$      &\ $ 3.689\pm0.296 $ &\ $104.859\pm0.851$ \\
$n=2(NP)$  & \  $2.840\pm0.191$ & \ $110.119\pm0.550$  & \ $2.963\pm0.188$  & \ $112.225\pm0.541$  \\
$n=3(NP)$ &\ $3.525\pm0.278$ & \ $113.308\pm0.800$       &\ $3.609\pm0.283$ &\ $115.238\pm0.813$   \\
$n=4(NP)$  & \  $3.624\pm0.141$ & \ $114.889\pm0.405$  & \ $3.741\pm0.121$  & \ $117.054\pm0.348$   \\\hline
$n=1(UP) $  & \  $3.147\pm0.112$ & \ $107.354\pm0.255$  & \ $3.277\pm0.119$  & \ $109.398\pm0.273$   \\
$n=2(UP) $  & \  $2.623\pm0.084$ & \ $113.337\pm0.193$  & \ $2.738\pm0.084$  & \ $115.603\pm0.193$  \\
$n=3(UP)$  & \  $3.126\pm0.118$ & \ $117.700\pm0.270$  & \ $3.189\pm0.126$  & \ $119.765\pm0.289$   \\
$n=4(UP)$  & \  $3.604\pm0.079$ & \ $118.370\pm0.180$  & \ $3.749\pm0.067$  & \ $120.571\pm0.153$   \\
\end{tabular}
\end{ruledtabular}
\end{table*}

\subsection*{3.4 Shell structure of the mass spectra }
 To get a clear outline of the baryon spectra, the shell structures of the mass spectra of $\Xi_{bb}$ and $\Omega_{bb}$ baryons are presented respectively.
    As shown in Figs.7 and 8, there lies a big gap (about 230 MeV) between the $1S$ and $1P$ sub-shells. This implies the experimental measurement of the $1S$ states for the $\Xi_{bb}$ or $\Omega_{bb}$ baryons could be done cleanly.

For each ground state, we average the predicted masses including ours and those listed in Table I, and compute the deviations (errors) from the mean value. The errors for different ground states are displayed in Fig.9. The mean values are obtained with $\bar{m}=\sum_{i=1}^{N}m_{i}/N$. The errors are calculated with $\Delta m_{i}=m_{i}-\bar{m}$ and listed in chronological order. There are a total of 43 theoretical results here.
  The obtained mean masses are as follows: $\bar{m}_{th}(\Xi_{bb})$=10183 MeV, $\bar{m}_{th}(\Xi_{bb}^{*})$=10234 MeV, $\bar{m}_{th}(\Omega_{bb})$=10287 MeV and $\bar{m}_{th}(\Omega_{bb}^{*})$=10337 MeV.

   As mentioned in section I, references~\cite{art013,art017,art018,art0182,art0183,art09,art025,art031,art056} have studied the mass spectral structures.  We compare our mass spectral structures with those of references~\cite{art0183} and~\cite{art056}. As shown in Table V, one can see the predicted masses of the $S$ states from the two references and ours are generally consistent with each other.
   Nevertheless, the difference between them starts with the $P$ states.
   In our calculations, there are only two $P$ states due to the limitation of the $\rho$-mode. But, references~\cite{art056} and~\cite{art0183} predict more $P$ states than ours. In Table V, we list some predicted mass values of the $P$ states from references~\cite{art056} and~\cite{art0183}, and label them in the way of $nL(J^{P})$.
  For these $P$ states, the masses from the reference~\cite{art056} are smaller than those from the reference~\cite{art0183} and ours as a whole. In addition, for the reference~\cite{art0183}, the predicted masses of the $P(\frac{1}{2}^{-})$ states are greater than those of the $P(\frac{3}{2}^{-})$ states, and the mass splittings of the $P$ doublets are rather small. In general, the mass spectral structures given by the three papers are different.

    In fact, it can be seen that different theoretical models speculate different inner structures of baryons and lead to different results. This reflects the complexity of theoretical calculation for doubly heavy baryons. We look forward to experimental testing of these models and further understanding of the structure of double bottom baryons.

\begin{figure}[htbp]
\begin{center}
\includegraphics[width=0.8\textwidth]{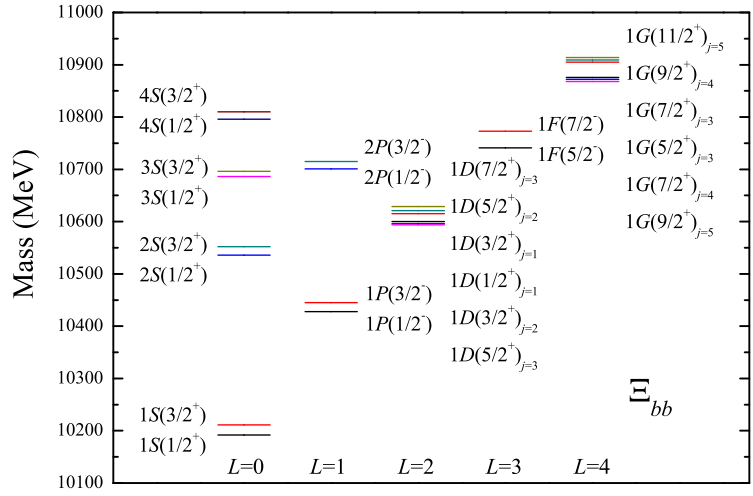}
\end{center}
\caption{(Color online)Shell structure of the $\Xi_{bb}$ family. The mass is measured in MeV.}
\end{figure}

\begin{figure}[htbp]
\begin{center}
\includegraphics[width=0.8\textwidth]{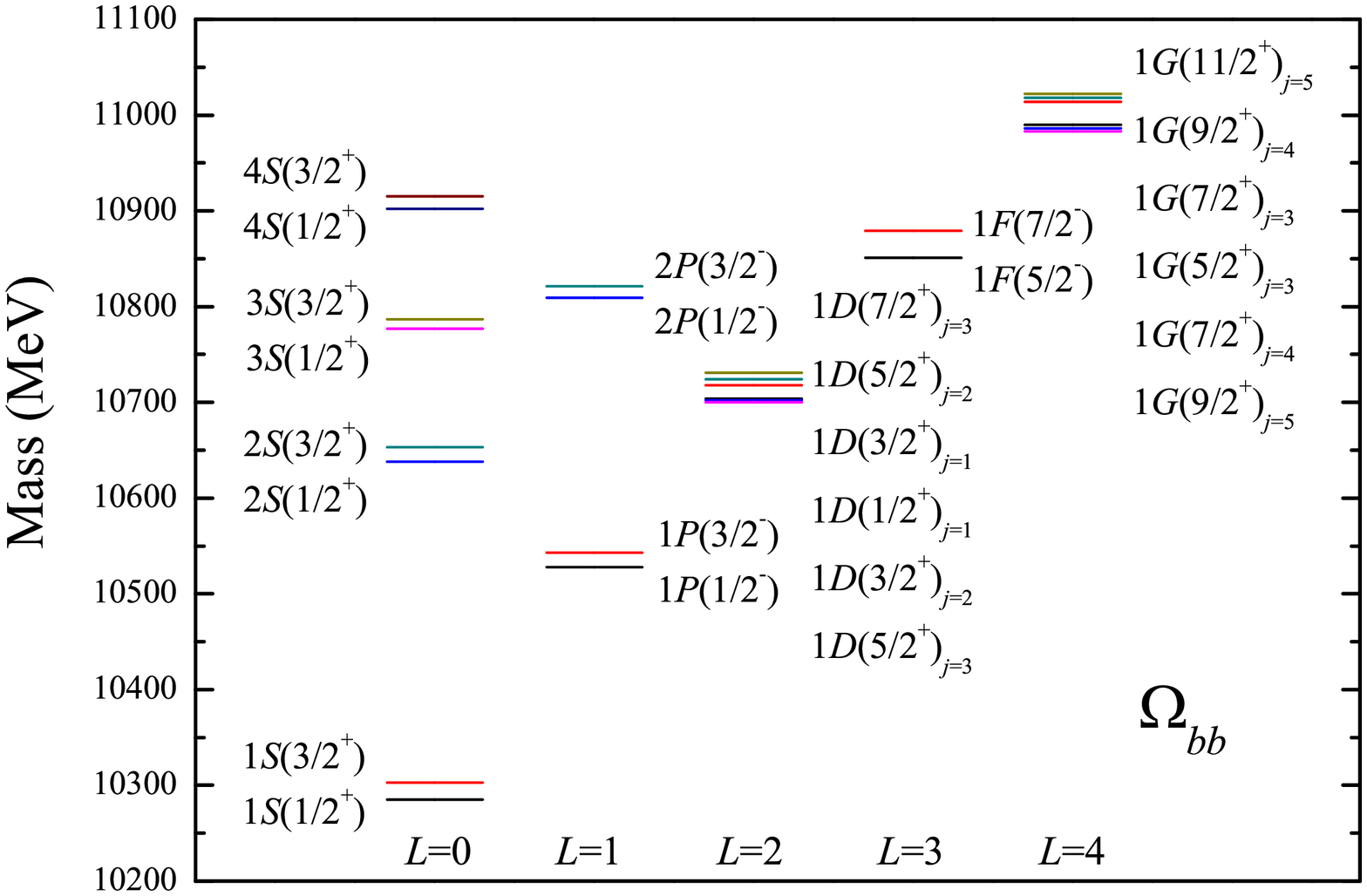}
\end{center}
\caption{(Color online)Same as Fig.7, but for the $\Omega_{bb}$ family.}
\end{figure}

\begin{figure}[htbp]
\begin{center}
\includegraphics[width=1.0\textwidth]{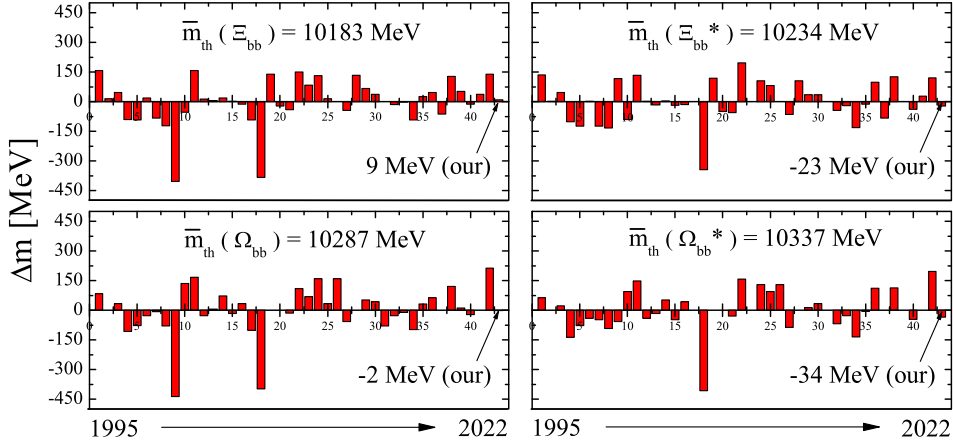}
\end{center}
\caption{(Color online)Mass errors from the mean masses of the theoretical calculated masses of the ground states for double bottom baryons given by the references in recent 30 years. The mass errors in this work are indicated. The mean masses are also presented. }
\end{figure}

\begin{table*}[htbp]
\begin{ruledtabular}\caption{The masses (MeV) of the $S$ and $P$ states for $\Xi_{bb}$ baryons.}
\begin{tabular}{c |c c c c| c c c c}
      & \ $n=1$ & \ $n=2$ & \ $n=3$ & \ $n=4$ & \ $n=1$ & \ $n=2$ & \ $n=3$ & \ $n=4$   \\ \hline
  $L(J^{P})$   & \multicolumn{4}{c|}{$S(\frac{1}{2}^{+})$}   &\multicolumn{4}{c}{$S(\frac{3}{2}^{+})$}   \\
     \hline
~\cite{art056}&\ 10202 &\ 10441   &\ 10630   &\ 10832  &\ 10237  &\ 10482 &\ 10673  &\ 10860 \\
 ~\cite{art0183}&\ 10221 &\ 10525   &\ 10749   &\ 10940  &\ 10261  &\ 10540 &\ 10756  &\ 10943 \\
             our  &\ 10192 &\ 10536    &\ 10686   &\ 10796  &\ 10211  &\ 10552 &\ 10696  &\ 10810 \\ \hline
  $L(J^{P})$    & \multicolumn{4}{c|}{$P(\frac{1}{2}^{-})$}   &\multicolumn{4}{c}{$P(\frac{3}{2}^{-})$}   \\
     \hline
~\cite{art056}&\ 10368 &\ 10563   &\ 10632   &\ 10744  &\ 10408  &\ 10607 &\ 10673  &\ 10788 \\
 ~\cite{art0183}&\ 10458 &\ 10686   &\ 10883   &\ 11055  &\ 10456  &\ 10685 &\ 10882  &\ 11055 \\
             our  &\ 10428 &\ 10701    &\ 10912   &\ 10959  &\ 10445  &\ 10715 &\ 10922  &\ 10973 \\

\end{tabular}
\end{ruledtabular}
\end{table*}

\section*{IV. Conclusions}

In this work, combining the relativistic quark model with the ISG method, we investigate the double bottom  baryon spectra systematically. The calculated result shows that the $\rho$-mode appears lower in energy than the other modes, which confirms the conjecture of the heavy quark excitation domination for the singly- and doubly-heavy baryons. Consequently, as a reasonable approximation, we obtain the mass spectra of the $\Xi_{bb}$ and $\Omega_{bb}$ families in the $\rho$-mode.
We also investigate the root mean square radii and the quark radial probability density distributions of these states, which deepens the understanding of the structure of double bottom baryons.

Based on the predicted mass spectra, we construct successfully the Regge trajectories in the $(J,M^{2})$ plane. These trajectories are not parallel to each other and not equidistant as a whole. Nevertheless, we can not currently construct the linear trajectories in the $(n,M^{2})$ plane, which is an apparent difference between our mass spectra and those obtained from some other models.

At last, the mass spectral structures of the $\Xi_{bb}$ and $\Omega_{bb}$ families are presented, from which we could get a bird's-eye view of the mass spectra. Then, we discuss the masses of the ground states of double bottom baryons and examine the reliability of our mass prediction through an extensive comparison with those given by the theoretical studies in the past three decades. We also analyze the differences between the structures of our spectra and those from other theoretical methods.

This work is a continuation of the series of our studies on the heavy baryons. The calculations in this work are systematic and the predicted masses and mass spectral structures might be a useful reference for related experiments.

\section*{Acknowledgements}

 This research is supported by the Central Government Guidance Funds for Local Scientific and Technological Development of China (No. Guike ZY22096024), the National Natural Science Foundation of China (Grant No. 11675265), the Continuous Basic Scientific Research Project (Grant No. WDJC-2019-13), and the Leading Innovation Project (Grant No. LC 192209000701).

\end{document}